\documentclass[slac_one,amsmath,nofootinbib]{revtex4}
\usepackage{graphicx,axodraw}
\usepackage{fancyhdr}
\def\cO#1{{\cal{O}}\left(#1\right)}

\newcommand{\beq}{\begin{equation}}
\newcommand{\eeq}{\end{equation}}
\newcommand{\beqa}{\begin{eqnarray}}
\newcommand{\eeqa}{\end{eqnarray}}

\begin{document}

\title{
%{\small{2005 ALCPG \& ILC Workshops - Snowmass,
%      U.S.A.}}\\
%  \vspace{12pt} 
Unstable particle production near threshold with
  effective theory methods }

\author{G.~Zanderighi} \affiliation{CERN, TH Division, CH-1211, Geneva
  23, Switzerland\\
 Fermilab, P.O. Box 500, 60510 Batavia, IL, US
}
\begin{abstract}
  We illustrate the use of effective theory methods to describe
  resonant unstable particles. We outline the necessary ingredients to
  describe $W$-pair production close to threshold in $e^-e^+$
  collisions.
\end{abstract}
\maketitle
\thispagestyle{fancy}
\section{INTRODUCTION} 
Physical processes at ongoing and upcoming colliders involve the
production and subsequent decay of heavy, unstable particles ($Z_0$,
$W^{+/-}$, $t$, \dots).
Their resonant production allows a precise determination of particle
properties. 
However, weak-coupling perturbation theory (PT) is known to fail in
the resonant region~\cite{Veltman:th}. The break-down of ordinary PT
is due to the appearance of a second small parameter, besides the
coupling constant: the width $\Gamma$ of the unstable particle in unit
of its mass $M$.
A self-energy resummation allows one to take into account finite width
effects, however this procedure introduces some arbitrariness often
reflected in gauge-dependent results.  Additionally, there is no clear
prescription on how to improuve systematically the accuracy of the
results.

We start from the observation that the presence of two small
parameters is {\it the} characteristic feature of the problem, so that
in a theory which formulates correctly the double expansion in the
coupling and in $\Gamma/M$ all other issues (gauge invariance,
resummation) should follow automatically.
As is typical in multi-scale problems, we use effective theory methods
to formulated this expansion.
This provides us with a computational scheme for performing
calculations of resonant production of unstable particles at any accuracy.

\section{$W$-pair production close to threshold}
In~\cite{BCSZ} we presented how an effective theory approach allows
one to describe processes with resonant unstable particles in
intermediate stages. We applied the method to the description of the
resonant production of a scalar, heavy particle.  This simple
toy-model allowed us to study the problem of treating consistently
finite width effects while keeping all technical difficulties to a
minimum.

Here we consider $W$-pair production close to threshold at an $e^+e^-$
collider. Elements of the calculation were first presented
in~\cite{Beneke:2004xd}.
We consider specifically the process
\begin{equation}
e^-(p_1)\, e^+(p_2) \to W^+(k_1)\, W^-(k_2) \to 
\mu^-(l_1)\, \bar{\nu}_\mu(l_2)\, u(l_3)\, \bar{d}(l_4)\,.
\label{wwprocess}
\end{equation}
At threshold $k_{1/2} \sim \{M_W(1 +v^2), \pm M_W\vec v\} $ and the
counting is $\alpha_{ew} \sim \alpha_s^2 \sim v^2\>$. 
This process is crucial for the precise determination of the $W$ mass
and a lot of work has been carried out in the last years to improuve
the accuracy of the description of $W^+W^-$ mediated four-fermion
final states.  Recently the full $\cO{\alpha_{ew}}$ to $e^+e^-$ to
four-fermion has been completed~\cite{Denner:2005es}.
Here we focus on obtaining results that are valid near threshold,
where ordinary PT and the double pole approximation break down.

The first step in the construction of the effective theory is to
integrate out hard fluctuations $k\sim M$ (so called factorizable
corrections), for which there is no quantum-interference with
resonant, slowly propagating particles.
This gives rise to the hard matching coefficients of an effective
theory where only soft, collinear or resonant modes are dynamical. 
The hard matching coefficients are determined by matching onshell
Green functions in full theory to operators in the effective theory.
Notice that the precise definition of what are the ``hard'' modes
depends on the process and on the observable under consideration.
The splitting between hard and dynamical does not involve a cut-off,
instead in dimensional regularization this splitting can be achieved
with the strategy of regions~\cite{Beneke:1998zp}. 

Once the matching coefficients and the effective operators have been
calculated at the required order in PT, calculations can be done
suitably in the effective theory framework.
Here we will outline how to organize the lowest order calculation and
the classes of terms contributing to the first order correction to it.
The aim is to make transparent how the calculation can be
systematically extended to higher orders.

\subsection{Leading order}
\begin{figure}[h]
\begin{center}
\begin{picture}(150,80)(0,0)
\SetScale{0.7}
\ArrowLine(40,40)(10,10)
\ArrowLine(10,70)(40,40)
\Text(5,10)[r]{$p_2$}
\Text(5,50)[r]{$p_1$}
\GCirc(40,40){2}{0.1}
\Photon(38,40)(100,67){-2}{6.5}
\Photon(38,40)(100,13){2}{6.5}
\Text(50,47)[c]{$k_1$}
\Text(50,10)[c]{$k_2$}
\GCirc(100,65){2}{0.1}
\GCirc(100,15){2}{0.1}
\ArrowLine(100,65)(136,80)
\ArrowLine(136,50)(100,65)
\ArrowLine(100,15)(136,0)
\ArrowLine(136,30)(100,15)
\Text(100,60)[l]{$l_1$}
\Text(100,36)[l]{$l_2$}
\Text(100,0)[l]{$l_3$}
\Text(100,22)[l]{$l_4$}
\end{picture}
\end{center}
\vspace{-.5cm} {\caption{\it Leading order Feynman diagram in the
effective theory.
\label{fig:LO}}}
\vspace{-0.4cm}
\end{figure}
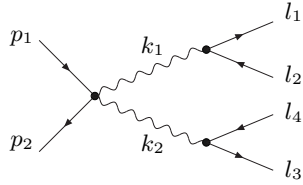

The leading order (LO) contribution is obtained by tree-level matching
of the on-shell operators and by resuming onshell one-loop
self-energies in the propagators.
Specifically, at LO one needs
\begin{itemize}
\vspace{-.1cm}
\item tree level matching for the production vertex $e^+e^-\to W^+W^-$ {
  $$
  {\cal L}^{(0)}_{\cal P} = \frac{2\pi
    \alpha_{ew}}{M_W^2}\left(\bar e_L \gamma^{\left[i\right.} i
    D^{\left.j\right]}
    e_L\right)\left(\Omega_-^{*i}\Omega_+^{*j}\right)\>,
  $$} 
where $\Omega_\pm$ denote the non-relativistic vector fields with mass
dimension 3/2 for the $W^\pm$ bosons.
Notice that at LO only the $e^-_Le^+_R$ amplitude contributes;
\vspace{-.2cm}
\item resummation of $\cO{\alpha_{ew}}$ onshell self-energies in the
  propagators
  $$
  {\cal L}^{(0)}_{\cal NR} = \sum_{\pm} \Omega_{\pm}^{*i} \left( i
    D^0+\frac{\vec
      D^2}{2M_W}-\frac{\Delta_1}{2}\right)\Omega_{\mp}^{i}\>,
  $$
  here $\Delta \equiv(s-M_W^2)/M_W$ is the hard matching
  coefficient and $\Delta_1$ denotes the leading contribution
  $\cO{M_W\alpha_{ew}}$.
  The W bosons are then described by a non-relativistic Lagrangian
  similar to NRQCD; 
\vspace{-.2cm}
\item tree level matching for the decay vertices  $W\to l \bar l$
  $$
  {\cal L}^{(0)}_{\cal D} = -\frac{g_{ew}}{\sqrt{2}} \Omega_{-}^{i}
  \bar{\mu}_L\gamma^i\nu_L -\frac{g_{ew}}{\sqrt{2}} \Omega_{+}^{i}
  \bar{u}_L\gamma^i d_L\>.
$$
\end{itemize}
To obtain the LO amplitude, one derives the Feynman rules
corresponding to these effective operators and combines the various
elements. The diagram contributing at LO in the effective theory is
shown in Fig.~\ref{fig:LO}.

\vspace{-.6cm}
%%%%%%%%%%%%%%%%%%%%%%%%%%%%%%%%%%%%%%%%%%%%%%%%%%%%%%%%%%%%%%%%%%%
%%%%%%%%%%%%%%%%%%%%%%%%%%%%%%%%%%%%%%%%%%%%%%%%%%%%%%%%

\subsection{$\cO{\alpha_s,v,\alpha_{ew}/v}$ corrections}
The first correction to the LO amplitude is obtained by including all
corrections $\cO{\alpha_s \sim v \sim \alpha_{ew}/v}$ to it.  Since
$\alpha_s \sim \alpha_{ew}^{1/2}$ we call this perturbative order
$N^{1/2}LO$.  The set of terms needed at this order is
\begin{itemize} 
\vspace{-.2cm}
\item $v$ corrections to production vertex 
\begin{eqnarray}
{\cal L}_p^{(1/2)} &=& \frac{c_1}{M_W^3}
\left(\bar{e}_L \gamma^{j} e_L \right) 
\left(\Omega_-^{*i} (-i) D^j \Omega_+^{*i}\right)
\label{LPhalf}
+ \frac{c_2}{M_W^3}
\left(\bar{e}_L \gamma^{[i} e_L \right) 
\left(\Omega_-^{*i} (-i) D^{j]} \Omega_+^{*j}\right) 
\nonumber \\
&+& \frac{c_3}{M_W^3}
\left(\bar{e}_L \gamma^{[i} n^{j]} n^l e_L \right) 
\left(\Omega_-^{*i} (-i) D^l \Omega_+^{*j}\right) 
+ \frac{c_4}{M_W^3}
\left(\bar{e}_L \gamma^j\gamma^l\gamma^i e_L \right) 
\left(\Omega_-^{*i} (-i)D^l \Omega_+^{*j}\right)\,,
\end{eqnarray}
with the matching coefficients
\begin{eqnarray}
c_1 = \pi\alpha_{ew}
\frac{M_Z^2 \sin^2\theta_w -2 M_W^2}{4 M_W^2-M_Z^2}\,; 
\qquad 
c_2 = \pi\alpha_{ew} 
\frac{M_Z^2 (1-2 \sin^2\theta_w)}{4 M_W^2-M_Z^2}\,; 
\qquad 
c_3 = 2 \pi\alpha_{ew} \,; 
\qquad 
c_4 = \pi\alpha_{ew}\,. 
\end{eqnarray}
Additionally, at $N^{1/2}LO$ there is a contribution from the $e^-_Re^+_L$ amplitude; 
\vspace{-.2cm}
\item two-loop $\alpha_s\alpha_{ew}$ corrections to the onshell
  propagator. These give rise to the matching coefficient
  $\Delta_{3/2}$, which can either be resummed, i.\ e.\ included in
  the effective operator, or included as perturbative interactions;
  \vspace{-.2cm}
\item $\alpha_s$ corrections to the decay stage, which cancel if one
  is inclusive on hadronic decay products. There are no $\cO{v}$
  corrections in the decay stage, the first non-trivial kinematical
  correction being $\cO{v^2}$;
  \vspace{-.2cm}
\item the exchange of one potential photon ($q_0 \sim M_Wv^2, \vec
  q \sim M_W v$), which is $\cO{\alpha_{ew}/v}$
\begin{equation}
{\cal A}^{(1/2,c)} = -i\, (4\pi\alpha)\, {\cal A}^{(0)} \times
\label{eq:Acoul}
\int \frac{d^D q}{(2\pi)^D}\, \frac{1}{\vec{q}^{\,2} }\,
\frac{1}{\left(E_1 - q^0 - \frac{(\vec{k}_1-\vec{q})^2}{2 M_W} 
      - \frac{\Delta^{(1)}}{2} + i\epsilon \right) } 
\frac{1}{\left(E_2 + q^0 - \frac{(\vec{k}_2+\vec{q})^2}{2 M_W} 
      - \frac{\Delta^{(1)}}{2} + i\epsilon\right) } ,
\end{equation}
where $\vec{k}\equiv\vec{k}_1=\vec{l}_1+\vec{l}_2$ and
$\vec{k}_2=\vec{l}_3+\vec{l}_4=-\vec{k}$.  
One obtains
\begin{equation}
{\cal A}^{(1/2,c)} = {\cal A}^{(0)} \, 
\frac{\alpha\, M_W}{|\vec{k}|} \times
\label{eq:AcoulF}
\arctan\frac{|\vec{k}|}
  {\sqrt{M_W(\Delta^{(1)}-E_1-E_2) - i\epsilon}},
\end{equation}
in agreement with~\cite{Fadin:1993kg}. 
The exchange of one potential photon turns out to be the only
contribution at $N^{1/2}LO$ which is not due to hard contributions.
~\footnote{Notice that in the similar $t\bar t$ threshold production
  case, a resummation of potential photons is necessary.}
\end{itemize} 
Combining these terms one obtains the $N^{1/2}LO$ amplitude in the
effective theory.
Similarly the calculation can be organized at higher orders. Higher
order corrections come from
\begin{itemize}
\vspace{-.2cm}
\item matching of lower order effective operators at higher accuracy in the
  expansion in the couplings;
\vspace{-.2cm}
\item kinematical corrections to the effective operators;
\vspace{-.2cm}
\item matching of higher order effective operators;
\vspace{-.2cm}
\item contribution from dynamical effective modes.
\end{itemize}
The power counting allows one to identify the terms needed at a given
order prior to performing the calculation and therefore makes a
systematic organization of the calculation straightforward.
The only bottleneck in going to higher orders remains the (standard)
complexity of multi-loop and multi-scale integrals, the same
difficulties one encounters in the treatment of stable particles.

\section{Conclusions}
We considered the resonant $W$-pair production with effective theory
methods. We outlined the calculation of the ${\cal O}(\alpha_s, v,
\alpha_{ew}/v)$ corrections. Work in progress is the calculation of
higher order corrections, though technically more difficult, no new
conceptual issue arise.

\begin{acknowledgments}
  I thank M.~Beneke, A.~Chapovsky, N.~Kauer and A.~Signer for
  collaboration in this project.
\end{acknowledgments}

%\fancyhead[C]{\it {XXII Texas Symposium on Relativistic Astrophysics at Stanford University,
\end{document}